\documentclass[aps,preprint,showpacs,preprintnumbers,eqsecnum,amsmath,amssymb]{revtex4}

\usepackage{graphics,epsfig}
\usepackage{graphicx}
\usepackage{dcolumn}
\usepackage{bm}

\begin{document}

\title{   Particle energy and Hawking temperature }
\author{Chikun Ding }
\author{Mengjie Wang}\author{ Jiliang Jing}
\thanks{Corresponding author, Electronic address:
jljing@hunnu.edu.cn} \affiliation{ Institute of Physics and
Department of Physics,
Hunan Normal University, Changsha, Hunan 410081, P. R. China \\
and
\\ Key Laboratory of Low Dimensional Quantum Structures and
Quantum Control of Ministry of Education, Hunan Normal University,
Changsha, Hunan 410081, P.R. China}

 \baselineskip=0.65 cm

\vspace*{0.2cm}
\begin{abstract}
\vspace*{0.2cm} Some authors have recently found that the tunneling
approach gives a different Hawking temperature for a Schwarzschild
black hole in a different coordinate system. In this paper, we find
that to work out the Hawking temperature in a different coordinate
system by the tunneling approach, we must use the correct definition
of the energy of the radiating particles. By using a new definition
of the particle energy, we obtain the correct Hawking temperature
for a Schwarzschild black hole in two dynamic coordinate systems,
the Kruskal-Szekers and dynamic Lemaitre coordinate systems.

\end{abstract}

 \vspace*{1.5cm}

\maketitle

\section{Introduction}

 In recent years, a semi-classical
method for controlling Hawking radiation as a tunneling effect has
been developed and has garnered much interest
\cite{ag,ding,ding1,wil,wil2,wil3,wil4,sh3,sh1,sh2,landau,man,
man1,man3,akh1,akh2,chow,par,par1,par2,
arz,jiang,zz,sqw,ajm,pm1,pm,pill1,pill2,sang1,chende,sean}. Angheben
{\it et al} \cite{ag} and Padmanabhan {\it et al} \cite{sh1,sh2,sh3}
used the complex path analysis that was developed by Mann {\it et
al} \cite{man,man1}. In this method, the semiclassical propagator
$K(r_2,t_2;r_1,t_1)$ in $(1+1)$ dimensional Schwarzschild spacetime
is $
K(r_2,t_2;r_1,t_1)=N\exp\big(\frac{i}{\hbar}I(r_2,t_2;r_1,t_1)\big),
$ where $I$ is the classical action of the trajectory to leading
order in $\hbar$ for a massless particle to propagate from
$(t_1,r_1)$ to $(t_2,r_2)$ and is constructed by using the
Hamilton-Jacobi frame. $N$ is a suitable normalization constant. We
can separate variables $I=-Et+W(r)$ due to the symmetries of the
spacetime, where $E$ is the energy of particle. This action acquires
a singularity at the event horizon in analogy with the quantum
tunneling process in quantum mechanics. In semiclassical quantum
mechanics, this singularity is regularized by specifying a suitable
complex contour \cite{landau}. In the case of a black hole
\cite{sh3}, we should take the contour to be an infinitesimal
semicircle above the pole $r=r_H$ for outgoing particles
($\partial_r I>0$) on the left of the horizon and ingoing particles
($\partial_r I<0$) on the right; similarly, for the ingoing
particles on the left and outgoing particles on the right of the
horizon (corresponding to the time reversed situation), the contour
is below the pole. After integrating around the pole, we find that
the action $I(r_2,t_2;r_1,t_1)$ is complex, so the probability
$\Gamma\propto e^{-2\text{Im}I}$ and the probability of the emission
of particles are not the same as the probability of absorption, the
ratio is $ \Gamma[\text{emission}]=e^{-8\pi
ME}\Gamma[\text{absorption}].$ This result shows that it is more
likely for a particular region to gain particles than lose them.
Further, the exponential dependence on the energy allows one to give
a `thermal' interpretation to this result. In a system with a
temperature $T_H$, the absorption and the emission probabilities are
related by
\begin{eqnarray}\label{gamma00}\Gamma[\text{emission}]=e^{-
E/T_H}\Gamma[\text{absorption}].\end{eqnarray} The above relation
can be interpreted to be equivalent to a thermal distribution of
particles in analogy with that observed in any system interacting
with black body radiation. Then, the standard Hawking temperature is
recovered.

However, in the case of a black hole, the action for ingoing
particles should be real, so we employ a normalization condition on
the action $I'=I(r_2,t_2;r_1,t_1)+K$, where $K$ can be a complex
constant that ensures that the imaginary of action for ingoing
particles is equal to zero. Thus, the probabilities are
\begin{eqnarray}\label{gamma1}
\Gamma[\text{emission}]\propto e^{-2\text{Im}[I_++K]},~~
 \Gamma[\text{absorption}]\propto
e^{-2\text{Im}[I_-+K]}=1,\end{eqnarray}the ratio is
\begin{eqnarray}\label{gamma2}
\Gamma[\text{emission}]=e^{-2[\text{Im}I_+-\text{Im}I_-]}
\Gamma[\text{absorption}],\end{eqnarray} then
\begin{eqnarray}\label{gamma0}
 e^{-E/T_H}=e^{-2[\text{Im}I_+-\text{Im}I_-]},\end{eqnarray}
where $I_\pm$ are the square roots of the relativistic
Hamilton-Jacobi equation (\ref{hj}) corresponding to outgoing and
ingoing particles.

Using this method, some authors recently found that the tunneling
approach gives a different Hawking temperature of the Schwarzschild
black hole using different coordinates. These coordinates are all
stationary metrics; but what about non-stationary metrics? If one
employs non-stationary metrics, e.g. the Kruskal-Szekers coordinates
or dynamic Lemaitre coordinates, some other amazing facts come to
light. For example, in Ref. \cite{sh2}, the authors pointed out that
``In the case of Kruskal coordinate [sic], which is the maximal
extension of Schwarzschild spacetime, it is easy to show that the
semiclassical action when expressed in terms of Kruskal coordinates
{\it does not contain the singularity}. (The HJ equation (of a
massless particle) when expressed in terms of the Kruskal
coordinates $(V,U, \theta, \varphi)$ is of the form $(\partial
S_0/\partial V)^2-(\partial S_0/\partial U)^2=0$ (for S-wave, i.e.
$l$ =0). The solution of the equation can be easily obtained and is
given by $S_0(V_2,U_2; V_1,U_1) = S_0(2, 1) = -p_V (V_2 - V_1)\pm
p_U(U_2- U_1).)$" That is to say, the Hawking temperature cannot be
recovered! As for dynamic Lemaitre coordinates, the authors
\cite{sh2} used the transformations $U=3(R^*-\tau^*)/4M$ and
$V=3(R^*+\tau^*)/4M$, which already change the primitivity of this
coordinate system. It is easy to see that these transformations will
gives the line element
\begin{eqnarray}\label{sluv}ds^2&=&\frac{4}{9}M^2\left[(1-U^{-2/3})(dV^2+dU^2)
-2(1+U^{-2/3})dVdU\right]\nonumber\\
\quad\quad\quad&&+4M^2U^{4/3}(d\theta^2+\sin^2\theta d\varphi^2)
    .\end{eqnarray}
Obviously, in the outer region $R^*-\tau^*>4M/3$, the metrics are
$g_{VV}=g_{UU}<0$, whereas in the inner region $R^*-\tau^*<4M/3$,
the metrics are $g_{VV}=g_{UU}>0$. That is, the time-like/space-like
character of $V$ or $U$ are reversed again when  they cross the
horizon. In Ref. \cite{man1}, the authors studied the Dirac particle
radiation in the Kruskal coordinates by mathematically setting
$\partial_\chi=N(X\partial_T+T\partial_X)$ and $\partial_\chi I=-E$.
Yet, its physical meaning was not specified.

How can we obtain the correct Hawking temperature of a black hole in
different coordinates? We learn from the formulism (\ref{gamma00})
that if the energy of the particles $E$ is incorrect, we cannot find
the correct Hawking temperature. Therefore, first of all, we should
clarify the energy of the radiating particles in different
coordinates. In this manuscript, we will study the problem
carefully.

The paper is organized as follows. In Sec. II, the different
coordinate representations for the Schwarzschild black hole are
presented. In Sec. III, the expression of the particle energy is
presented. In Sec. IV, the Hawking temperature of the Schwarzschild
black hole from scalar particle tunneling in Kruskal-Szekers
coordinates is investigated. In Sec. V, the Hawking temperature of
the Schwarzschild black hole from scalar particle tunneling in
dynamic Lemaitre coordinates is studied. The last section is devoted
to a summary.

 \vspace*{0.4cm}
\section{Coordinate representations for a Schwarzschild black hole}

In standard coordinates, the line element of Schwarzschild black
hole is\begin{eqnarray}\label{ss}
ds^2=-(1-\frac{2M}{r})dt^2+\frac{1}{1-\frac{2M}{r}}dr^2+r^2d\theta^2
+r^2\sin^2\theta d\varphi^2,
   \end{eqnarray}
with an event horizon $r_H=2M$. We introduce two different
coordinate representations for the static black hole below.

\subsection{Kruskal-Szekers coordinate representation}
 The Kruskal-Szekers coordinate transformation is\begin{eqnarray}
 \label{skt}
&&\text{when
}r>2M,~\tau=\sqrt{\frac{r}{2M}-1}~e^{r/4M}\sinh(\frac{t}{4M}),
R=\sqrt{\frac{r}{2M}-1}~e^{r/4M}\cosh(\frac{t}{4M}),\nonumber\\&&
\text{when
}r<2M,~\tau=\sqrt{1-\frac{r}{2M}}~e^{r/4M}\cosh(\frac{t}{4M}),
R=\sqrt{1-\frac{r}{2M}}~e^{r/4M}\sinh(\frac{t}{4M}).
   \end{eqnarray}
The line element (\ref{ss}) in four dimensional spacetime becomes
\begin{eqnarray}\label{sk}
&&ds^2=\frac{32M^3}{r}~e^{-\frac{r}{2M}}(-d\tau^{2}+dR^{2})+r^2
(d\theta^2+\sin^2\theta d\varphi^2),
   \end{eqnarray}where $R=\tau$ at the event horizon $r_H=2M$. It is
   easy to see its metric $g_{\mu\nu}$ is the function of $\tau$,
   $R$ and $\theta$, so it is a dynamic coordinate system. In this
   coordinate, its coordinate singularity has been removed.
   $\tau$ is a time and $R$ is a space coordinate inside and
   outside the horizon.

\subsection{Dynamic
Lemaitre coordinate representation}

The dynamic Lemaitre coordinate representation is \cite{zel}
\begin{eqnarray}\label{sl}
&&ds^2=-d\tau^{*2}+\left[\frac{3}{4M}(R^*-\tau^*)\right]^{-2/3}dR^{*2}\nonumber\\
&&\quad\quad\quad+(2M)^2 \left[\frac{3}{4M}(R^*-\tau^*)\right]^{4/3}
(d\theta^2+\sin^2\theta d\varphi^2),
   \end{eqnarray}
where $R^*-\tau^*=4M/3$ at the event horizon $r_H=2M$. In this
coordinate system, the coordinate singularity has also been removed.
The proper time is equal to coordinate time, the $R^*$-axis is a
spatial axis and the $\tau^*$-axis is a temporal one not only inside
but also outside the event horizon. The geodesic of the particles is
continuous at the horizon.

 \vspace*{0.4cm}

\section{The definition of the energy of radiating particles}

     According to
quantum fields in curved spacetime, it is well known that one can
define the particle energy as long as the spacetime has
temporal-translational invariance and this energy is conserved. In
standard coordinates, line element (\ref{ss}) obviously has
temporal-translational invariance, so the particle energy is
$E=-\partial_t I$, where $I$ is the particle action, which can be
found via separating variables $I=-Et+I'(\vec{x})$. However, the
line elements (\ref{sk}) and (\ref{sl}) do not have this
temporal-translational Invariance; therefore, $\partial_\tau I$ is
not a constant of the motion. However, the particle energy should be
a conserved quantity, so the key problem is how to find the
expression of the particle energy in different coordinate systems.

As mentioned in \cite{hans}, for the particles moving along a
geodesic, the scalar product between the time-like Killing vector
and the particle four-momentum $p^\mu=mdx^\mu/d\lambda$ is a
constant, i.e.
\begin{eqnarray}\label{pp3}\xi_\mu p^\mu=\text{constant}.\end{eqnarray}
Furthermore, this quantity  $\xi_\mu p^\mu$ is not only a conserved
quantity along the geodesic, but also an invariant quantity in
different coordinates.  In a word, this quantity is a good one and
we can use it to define the particle energy in different coordinate
representations.

Consider the following Lagrangian of the massive radiating particle.
\begin{eqnarray}\label{lag}L=\frac{1}{2}mg_{\mu\nu}\frac{dx^\mu}
{d\lambda}\frac{dx^\nu}{d\lambda},\end{eqnarray} where $\lambda$ is
an affine parameter defined along the geodesic. Constructing the
action function $I=\int L d\lambda$, the possible physical  process
demands that, for variations $\delta I=0$, one can obtain the
Euler-Lagrangian
equation\begin{eqnarray}\label{lageq}\frac{d}{d\lambda}\Big(\frac{\partial
L}{\partial \dot{x}^\mu}\Big)-\frac{\partial L}{\partial x^\mu}=0,
\end{eqnarray} where the overdot
represents derivation with respect to the affine parameter
$\lambda$. From the Euler-Lagrangian equation (\ref{lageq}), the
respective conjugate momentum $p_\mu$ is \begin{eqnarray}
p_\mu=\frac{\partial L}{\partial\dot{x}^\mu}=\int\frac{\partial
L}{\partial x^\mu}d\lambda=\frac{\partial }{\partial x^\mu}\int L
d\lambda=\partial_\mu I,\end{eqnarray} and the constant $\xi_\mu
p^\mu=\xi^\mu p_\mu=\xi^\mu\partial _\mu I$. In the standard
coordinate representation (\ref{ss}), the time-like Killing vector
is $\tilde{\xi}^\mu=(1,0,0,0)$, and the metric tensor is independent
of the time coordinate $t$. Uing the Lagrangian nomenclature, $t$ is
a cyclic coordinate, $p_t$ is the conjugate momentum, and the
particle energy $E$ is the projection of four-momentum on the
time-like tetrad. Thus, $E=-p_t=-\partial _tI$. For this case,
$E=-\partial _tI=-\tilde{\xi}^\mu\tilde{\partial}_\mu I
=-\tilde{\xi}^\mu \tilde{p}_\mu$, hence this constant can be defined
as the particle energy, i.e.
\begin{eqnarray}\label{energy0}E=-\xi^\mu p_\mu.\end{eqnarray} When the particles travel
 from the exterior
region to interior region, the Killing vector changes its character
into space-like, but the numerical value of $\xi^\mu p_\mu$ is still
conserved \cite{doran}.

In the Kruskal-Szekers coordinate system, using transformation
(\ref{skt}), the Killing vector is\begin{eqnarray}\label{kill2}
~\xi^\mu=\frac{\partial x^\mu}{\partial
\tilde{x}^\nu}\tilde{\xi}^\nu=\left(\frac{R}{4M},~\frac{\tau}{4M},~0,~0\right).
   \end{eqnarray}
   Then, the energy of test particle is
   \begin{eqnarray}\label{energy1}E=-\xi^\mu p_\mu=-\xi^\mu\partial_\mu I
   =-(\frac{R}{4M}\partial_{\tau}+\frac{\tau}{4M}\partial_{R})
   I.\end{eqnarray}
In the dynamic Lemaitre coordinates, the Killing vector is
\begin{eqnarray}\label{xi}
\xi^\mu=\frac{\partial x^\mu}{\partial \tilde{x}^\nu}\tilde{\xi}^\nu
  =(1,~1,~0,~0),
   \end{eqnarray}
and then the particle energy is
\begin{eqnarray}\label{energy}E=-\xi^\mu p_\mu=-\xi^\mu\partial_\mu I
    =-(\partial_{\tau^*}+\partial_{R^*})I.\end{eqnarray}
The significance of this definition is that it specifies our
conception of the particle energy in different coordinates. We will
see that the Hawking temperature can be obtained using this
expression of the energy.

 \vspace*{0.4cm}
\section{Temperature of Schwarzschild black hole in the
 Kruskal-Szekers coordinate representation}

In this section, we use the energy definition above in the study of
the Hawking temperature of the Schwarzschild spacetime by employing
the Kruskal-Szekers coordinates (\ref{sk}).

Applying the WKB approximation
\begin{eqnarray}\label{ans}
\phi(\tau,R,\theta,\varphi)=\exp\Big[\frac{i}{\hbar}I(\tau,R,\theta,
\varphi)+I_1(\tau,R,\theta,\varphi) +\mathcal{O}(\hbar)\Big]
   \end{eqnarray}
to the Klein-Gordon equation\begin{eqnarray}\label{kg}
\frac{1}{\sqrt{-g}}\partial_{\mu}\big[\sqrt{-g}g^{\mu\nu}
\partial_{\nu}\phi\big]
-\frac{m^2}{\hbar^2}\phi=0,
   \end{eqnarray}
 then, to leading order in $\hbar$, we obtain the
   following relativistic Hamilton-Jacobi equation
   \begin{eqnarray}\label{hj}
g^{\mu\nu}\partial_{\mu} I\partial_{\nu} I+m^2=0.
   \end{eqnarray}
Now the action $I$ is the Hamiltonian principal function with
canonical momentum $p_{\mu}=\partial_{\mu} I$, and the Hamiltonian
$H=\frac{1}{2m}g^{\mu\nu}p_{\mu}p_{\nu}
=\frac{1}{2m}g^{\mu\nu}(\tau,R,\theta)p_{\mu}p_{\nu}$, so the time
$\tau$ is not the coordinate that can be disregarded. The momentum
$p_\tau=\partial_\tau I$ is not a constant, and we cannot separate
$\tau$ from $R$ in the action $I$ as in the Refs.
\cite{pill1,pill2}. In this coordinate representation, there exists
a solution of the form
 \begin{eqnarray}\label{ansatz}
I=I_0(\tau,R)+ J(\theta,\varphi)+K.
   \end{eqnarray}
   Inserting Eq. (\ref{ansatz}) and the metric (\ref{sk}) into the
Hamilton-Jacobi equation (\ref{hj}), we obtain a partial
differential equation
   \begin{eqnarray}\label{separation1}
\frac{r}{32M^3}~e^{\frac{r}{2M}}\left[-(\partial_{\tau}
I_0)^2+(\partial_{R} I_0)^2\right]+g^{ij}J_iJ_j+m^2=0.
   \end{eqnarray}
Using Eq. (\ref{energy1}), substituting $\partial_{\tau}
I_0=-\frac{4M}{R}(E+\frac{\tau}
   {4M}\partial_{R}I_0)$ into Eq.
   (\ref{separation1}), we obtain
\begin{eqnarray}\label{ww2}
(\partial_{R}I_0)_{\pm}=\frac{4ME\tau\pm
R\sqrt{16M^2E^2-\left[R^{2}-\tau^{2}\right]\frac{32M^3}{r}
~e^{-\frac{r}{2M}} (g^{ij}J_iJ_j+m^2)}} {R^{2}-\tau^{2}},
   \end{eqnarray}where $i,j=\theta,\varphi;~J_i=\partial_iI$. One solution of the Eq. (\ref{ww2})
corresponds to the scalar particles moving away from the black hole
(i.e. ``+" outgoing), and the other solution corresponds to
particles moving toward the black hole (i.e. ``-" incoming). To find
the relation between the total differential coefficient $dI_0$ and
the partial differential coefficients $\partial_{\tau} I_0$ or
$\partial_{R} I_0$, we need to know $\partial_{\tau} I_0$. From Eq.
(\ref{energy1}) and (\ref{ww2}) we obtain
\begin{eqnarray}
(\partial_{\tau}I_0)_{\pm}=-\frac{4MER\pm
\tau\sqrt{16M^2E^2-\left[R^{2}-\tau^{2}\right]\frac{32M^3}{r}
~e^{-\frac{r}{2M}} (g^{ij}J_iJ_j+m^2)}} {R^{2}-\tau^{2}}.
   \end{eqnarray}
   It is easy to prove
 \begin{eqnarray}\partial_{R}(\partial_{\tau}
I_0)=\partial_{\tau}(\partial_{R} I_0),~~dI_0=\partial_{R}
I_0dR+\partial_{\tau} I_0d\tau,\end{eqnarray}so the definite
integration of $I_0$ is
\begin{eqnarray}\label{I1}I_0&=&\int\partial_{R}
I_0dR+\partial_{\tau} I_0d\tau\nonumber\\&=& \int\partial_{R}
I_0\big(dR-\frac{\tau}{R}d\tau\big)-\int\frac{4ME}{R}
d\tau\nonumber\\&=&
\frac{1}{2}\int\frac{\partial_{R}I_0}{R}d\big(R^{2}-\tau^{2}
\big)-\int\frac{4ME}{R} d\tau .\end{eqnarray}
 Imaginary parts of the action can only come from the
pole at the horizon, so the second integration of Eq. (\ref{I1}) is
real, which shows that there is no temporal contribution in the
Kruskal-Szekers coordinate system. Integrating around the pole
$R=\tau$ at the horizon leads to
\begin{eqnarray}
 (\text{Im}I_0)_\pm&=&\text{Im}\left[\frac{1}{2}\int\frac{4ME\frac{\tau}
 {R}\pm
\sqrt{16M^2E^2-(R^{2}-\tau^{2})\frac{32M^3}{r}~e^{-\frac{r}{2M}}
(g^{ij}J_iJ_j+m^2)}}
{R^{2}-\tau^{2}}d(R^{2}-\tau^{2})\right],\nonumber\\
(\text{Im}I_0)_+&=&4\pi ME,~(\text{Im}I_0)_-=0.
   \end{eqnarray}
The probability of tunneling particles is
   \begin{eqnarray}\label{scalargamma}
\frac{\Gamma[\text{emission}]}{\Gamma[\text{absorption}]}=
\exp\left[-2(\text{Im}I_+-\text{Im}I_-)\right] =\exp\left[-8\pi
ME\right].
   \end{eqnarray}
Then, we obtain
   the Hawking temperature
\begin{eqnarray}\label{HT}
T_H=\frac{1}{8\pi M},
   \end{eqnarray}
which shows that the temperature of Schwarzschild black hole is the
same as that found in previous work using standard coordinates
\cite{sh2}.

 \vspace*{0.4cm}
\section{Hawking temperature of the Schwarzschild black hole in
The dynamic Lemaitre coordinate representation}

The definition of particle energy can also be used in the dynamic
Lemaitre coordinate representation. In this section, we study the
temperature of the Schwarzschild black hole in dynamic Lemaitre
coordinates (\ref{sl}) due to scalar particle tunneling.

We also cannot separate the time coordinate $\tau^*$ from the radial
coordinate $R^*$ in the form of the particle's action, so there
exists a solution in the form
   \begin{eqnarray}\label{ansatz2}
I=I_0(\tau^*,R^*)+ J(\theta,\varphi)+K.
   \end{eqnarray}
   Inserting the metric (\ref{sl}) and Eq. (\ref{ansatz2}) into
   the Hamilton-Jacobi
equation (\ref{hj}), we obtain
   \begin{eqnarray}\label{separation2}
-(\partial_{\tau^*}
I_0)^2+\left[\frac{3}{4M}(R^*-\tau^*)\right]^{2/3}(\partial_{R^*}
I_0)^2+g^{ij}J_iJ_j+m^2=0.
   \end{eqnarray}
Substituting Eq. (\ref{energy}) into Eq.
   (\ref{separation2}), we obtain
\begin{eqnarray}\label{ww}
&&(\partial_{R^*}I_0)_{\pm}=\frac{E\pm
\sqrt{\left[\frac{3}{4M}(R^*-\tau^*)\right]^{2/3}E^2-\{\left[\frac{3}{4M}
(R^*-\tau^*)\right]^{2/3}-1\}(g^{ij}J_iJ_j+m^2)}}
{\left[\frac{3}{4M}(R^*-\tau^*)\right]^{2/3}-1}.
   \end{eqnarray}
One solution of Eq. (\ref{ww}) corresponds to the scalar particles
moving away from the black hole (i.e. ``+" outgoing) and the other
solution corresponds to particles moving toward the black hole (i.e.
``-" incoming). In order to seek the relation between the total
differential $dI_0$ and partial differential
$\partial_{R^*}I_0,\partial_{\tau^*} I_0$, we need to find
$\partial_{\tau^*} I_0$. From Eq. (\ref{energy}) and (\ref{ww}) we
obtain
 \begin{eqnarray}(\partial_{\tau^*} I_0)_{\pm}=-\left\{E+\frac{E\pm
\sqrt{\left[\frac{3}{4M}(R^*-\tau^*)\right]^{2/3}E^2-\{\left[\frac{3}{4M}
(R^*-\tau^*)\right]^{2/3}-1\}(g^{ij}J_iJ_j+m^2)}}
{\left[\frac{3}{4M}(R^*-\tau^*)\right]^{2/3}-1}\right\}.\nonumber \\
\end{eqnarray} It is easy to prove
 \begin{eqnarray}\partial_{R^*}(\partial_{\tau^*}
I_0)=\partial_{\tau^*}(\partial_{R^*} I_0),~~dI_0=\partial_{R^*} I_0
dR^*+\partial_{\tau^*} I_0 d\tau^*,\end{eqnarray} therefore the
definite integration of $I_0$ is
 \begin{eqnarray}\label{I0}I_0&=&\int_{(R^*_0,~\tau^*_0)}^{(R^*_1,~\tau^*_1)}
 \Big[\partial_{R^*}
I_0dR^*+\partial_{\tau^*} I_0d\tau^*\Big]\nonumber\\&=&
\int_{(R^*_0,~\tau^*_0)}^{(R^*_1,~\tau^*_1)}\Big[\partial_{R^*}
I_0dR^*+(-E-\partial_{R^*}I_0)
d\tau^*\Big]\nonumber\\&=&\int_{(R^*_0-\tau^*_0)}^{(R^*_1-\tau^*_1)}\partial_{R^*}
I_0d(R^*-\tau^*)-\int_{\tau^*_0}^{\tau_1^*} Ed\tau^*,\end{eqnarray}
 where the point
$(R^*_0,\tau^*_0)$ is inside the event horizon
$\tau^*=R^*-\frac{4M}{3}$, and the point $(R^*_1,\tau^*_1)$ is
outside the horizon. Imaginary parts of the action can only come
from the pole at the horizon, so that the second integration of
(\ref{I0}) is real, which tells us that there is no temporal
contribution in the dynamic Lemaitre coordinate system. Substituting
Eq. (\ref{ww}) into (\ref{I0}), then integrating around the pole
$R^*-\tau^*=4M/3$ at the horizon leads to
\begin{eqnarray}\label{ww3}
 (\text{Im}I_0)_\pm&=&\text{Im}\left[\int\frac{E\pm
\sqrt{\left[\frac{3}{4M}(R^*-\tau^*)\right]^{2/3}E^2-\{\left[\frac{3}{4M}
(R^*-\tau^*)\right]^{2/3}-1\}(g^{ij}J_iJ_j+m^2)}}
{\left[\frac{3}{4M}(R^*-\tau^*)\right]^{2/3}-1}d(R^*-\tau^*)\right],
\nonumber\\(\text{Im}I_0)_+&=&4\pi ME,~(\text{Im}I_0)_-=0.
   \end{eqnarray}
The probability of a particle tunneling from inside to outside the
horizon is
   \begin{eqnarray}\label{scalargamma1}
\frac{\Gamma[\text{emission}]}{\Gamma[\text{absorbtion}]}=
\exp\left[-2(\text{Im}I_+-\text{Im}I_-)\right] =\exp\left[-8\pi
ME\right].
   \end{eqnarray}
We also obtain the correct Hawking temperature.

From the above discussions, it is easy to see that, with the
definition of the radiating particle energy (\ref{energy0}), the
problem of the Hawking radiation in Kruskal-Szekers and dynamic
Lemaitre coordinates is solved, and the Hawking temperature is
invariant.

 \vspace*{0.4cm}
\section{summary}

To study the Hawking radiation of a black hole in different
coordinates, we learn from the formulism (\ref{gamma00}) that the
key step is to define the energy of the radiating particles in
different coordinates. By means of the Euler-Lagrangian equation and
using the fact that $\xi_\mu p^\mu$ is a constant in coordinate
transformations, we present an expression of the energy of the
radiating particles: $E=-\xi^\mu p_\mu$.

As examples, we study the Hawking temperature of the Schwarzschild
black hole in the Kruskal-Szekers and dynamic Lemaitre coordinates
using the definition of the energy of the particles. In these two
coordinates, there are no coordinate singularities at the event
horizon, and there is no inversion between time and space across the
event horizon. We find that the Hawking temperature is invariant
under these two dynamic coordinate representations.

 \vspace*{0.2cm}
\begin{acknowledgments}
This work was supported by the National Natural Science Foundation
of China under Grants No. 10675045 and 10875040, the FANEDD under
Grant No. 200317, the Hunan Provincial Natural Science Foundation of
China under Grant No. 08JJ3010, the Hunan Provincial Innovation
Foundation for Postgraduate, and the Project of Knowledge Innovation
Program (PKIP) of the Chinese Academy of Sciences under Grant No.
KJCX2.YW.W10.

\end{acknowledgments}


\vspace*{0.2cm}
\begin{thebibliography}{99}


\bibitem{ag} M. Angheben, M. Nadalini,
L. Vanzo, and S. Zerbini, J. High Energy Phys. 0505 (2005) 014.

\bibitem{ding} C. K. Ding and J. L. Jing, Class. Quantum Grav. {\bf25} 145015 (2008).
\bibitem{ding1} C. K. Ding and J. L. Jing, Gen. Relat. Gravit. {\bf41} (to be
published) (2009).
\bibitem{wil}M. K. Parikh and F. Wilczek,
 Phys. Rev. Lett. {\bf85} 5042 (2000).
\bibitem{wil2} P. Kraus, and Frank Wilczek, Mod. Phys. Lett. A {\bf9}, 3713 (1994).
 \bibitem{wil3}P. Kraus and F.
Wilczek, Nucl. Phys. B {\bf433} 403 (1995).


 \bibitem{wil4}P. Kraus
and F. Wilczek, Nucl. Phys. B {\bf437} 231 (1995).
\bibitem{landau} L. D. Landau and E. M. Lifshitz, {\it Quantum Mechanics
(Non-relativistic Theory),} Course of Theoretical, Volume 2
(Pergamon, New York, 1975).

\bibitem{sh3} K. Srinivasan and T. Padmanabhan, Phys. Rev. D {\bf60} 024007 (1999).
\bibitem{sh1} S. Shankaranarayanan and K. Srinivasan and T. Padmanabhan, Mod. Phys. Lett. A
{\bf16} 571 (2001).

\bibitem{sh2}S. Shankaranarayanan and T. Padmanabhan and K. Srinivasan, Class. Quantum Grav. {\bf19} 2671 (2002).



\bibitem{man}R. Kerner and R. B. Mann, Phys. Rev.
D {\bf73} 104010 (2006).

\bibitem{man1}R. Kerner and R. B. Mann, Class. Quantum Grav. {\bf25} 095014 (2008).
\bibitem{man3}R. Kerner and R. B. Mann, Phys. Lett. B {\bf665} 277 (2008).
\bibitem{akh1}E. T. Akhmedov, V. Akhmedova and D. Singleton, Phys. Lett. B {\bf642} 124 (2006).
 \bibitem{akh2}E. T. Akhmedov, V. Akhmedova, D. Singleton and T.
 Pilling, Int. J. Mod. Phys. A {\bf22} 1705 (2007).
\bibitem{chow}B. D. Chowdhury, Pramana {\bf70} 593 (2008).




\bibitem{par} M. K. Parikh, Phys.
Lett. B {\bf546} 189 (2002).
\bibitem{par1}M. K. Parikh, Int. J.
Mod. Phys. D {\bf13} 2351 (2004).
\bibitem{par2}M. K. Parikh,
arXiv: hep-th/0402166.




\bibitem{arz} M. Arzano, A.
Medved and E. Vagenas, J. High Energy Phys. 0509 (2005) 037.

\bibitem{jiang} Qing-Quan Jiang,
Shuang-Qing Wu, and Xu Cai, Phys.Rev. D {\bf73} 064003 (2006).

\bibitem{zz} Jingyi Zhang, and Zheng Zhao, Phys.
Lett. B {\bf638} 110 (2006).

\bibitem{sqw} Shuang-Qing Wu, and
Qing-Quan Jiang, J. High Energy Phys. 0603 (2006) 079.



\bibitem{ajm} A. J. M. Medved and E. Vagenas, Mod. Phys. Lett. A {\bf20}
2449 (2005).

 \bibitem{pm1}B. Chatterjee, A. Ghosh and
P. Mitra, Phys. Lett. B {\bf661} 307 (2008).
\bibitem{pm} P. Mitra,
 Phys. Lett.
B {\bf648} 240 (2007).

\bibitem{pill1} V. Akhmedova, T. Pilling, A. de Gill and D. Singleton,
 Phys. Lett. B {\bf666}, 269 (2008).
\bibitem{pill2}E. T. Akhmedov, T. Pilling and D. Singleton,
 Int. J. Mod. Phys. D {\bf17} 2453 (2009).
\bibitem{sang1} S. P. Kim, J. High Energy Phys. 0711 (2007) 048.

\bibitem{chende} D. Y. Chen, Q. Q. Jiang and X. T. Zu,
Class. Quantum. Grav. {\bf25}, 205022 (2008).
\bibitem{sean} S. Stotyn, K. Schleich and D. Witt,
Class. Quantum. Grav. {\bf26}, 065010 (2009).
\bibitem{zel} Y. A. B. Zeldovich and I. D. Novikov, {\it Rilativistic
Astrophysics I}, Chicago: Univ of Chicgo Press, (1971).

\bibitem{hans}H. C. Ohanian and R. Rrffini, {\it GRAVITATION AND
SPACETIME (2nd ed.)}, W. W. Norton \& Company, Inc. (1994).
\bibitem{doran} R. Doran, F. S. N. Lobo and P. Crawford, Found.
Phys. {\bf 38}, 160 (2008).




\end{thebibliography}
\end{document}